\documentclass{aa}  

\usepackage{graphicx}
\usepackage{natbib}
\usepackage{txfonts}
\usepackage{color}
\providecommand{\lsol}{\ensuremath{L_{\odot}}}
\providecommand{\msol}{\ensuremath{M_{\odot}}}
\providecommand{\rsol}{\ensuremath{R_{\odot}}}
\providecommand{\teff}{\ensuremath{T_{\rm eff}}}

\providecommand{\bprpzero}{\ensuremath{(G_{\rm BP}-G_{\rm RP})_0}}
\providecommand{\bprp}{\ensuremath{(G_{\rm BP}-G_{\rm RP})}}
\providecommand{\ebpminrp}{\ensuremath{E(G_{\rm BP}-G_{\rm RP})}}
\providecommand{\ebv}{\ensuremath{E(B - V)}}

\providecommand{\feh}{\rm [Fe/H]}
\providecommand{\alfe}{[\ensuremath{\alpha}/{\rm Fe}]}
\providecommand{\numax}{\ensuremath{\nu_{\rm max}}}
\providecommand{\fbol}{\ensuremath{F_{\rm bol}}}
\providecommand{\mg}{\ensuremath{M_{\rm G}}}

\providecommand{\pmm}{\ensuremath{\pm}}

\begin{document}

   \title{Benchmarking the effective temperature scale of red giant branch stellar models: the case of the metal-poor halo giant HD~122563}

%   \subtitle{}

   \author{O.L.~Creevey\inst{1} 
   \and S.~Cassisi\inst{2,3}
   \and F.~Th{\'e}venin\inst{1}
   \and M.~Salaris\inst{2,4}
   \and A.~Pietrinferni\inst{2}
          }

   \institute{Universit{\'e} C{\^o}te d'Azur, Observatoire de la C{\^o}te d'Azur, CNRS, Laboratoire Lagrange, France \email{orlagh.creevey@oca.eu}
   \and
   INAF-Osservatorio Astronomico d'Abruzzo, via M. Maggini, sn.
     64100, Teramo, Italy \email{santi.cassisi@inaf.it}
     \and 
     INFN -  Sezione di Pisa, Largo Pontecorvo 3, 56127 Pisa, Italy
     \and
     Astrophysics Research Institute, Liverpool John Moores University, 146 Brownlow Hill, Liverpool, L3 5RF, UK
     }

   \date{Received ; accepted }

% \abstract{}{}{}{}{} 
% 5 {} token are mandatory

\abstract{There is plenty of evidence in the literature of significant discrepancies between the observations and models of metal-poor red giant branch stars, in particular regarding the effective temperature, \teff, scale.}
{We revisit the benchmark star HD\,122563 using the most recent observations from Gaia Data Release 3, to investigate if these new constraints may help in resolving this discrepancy. } 
{We review the most recent spectroscopic determinations of the metallicity of HD\,122563 [Fe/H], and provide a new assessment of its fundamental parameters, i.e. bolometric luminosity, \teff, surface gravity, plus 
a photometric determination of its metal content. 
Using these constraints, we compare the position of the star in the Hertzsprung-Russell (H-R) diagram with various recent sets of stellar evolution tracks.
}
{The H-R diagram analysis reveals a significant disagreement between 
observed and theoretical \teff\ values, 
when adopting the most recent spectroscopic estimate of [Fe/H]. 
On the other hand, by using the photometric determination of HD\,122563 [Fe/H] some of the selected sets of stellar tracks appear in fair agreement with observations. The sets with discrepant \teff\ can be 
made to agree with observations 
either by modifying the prescription adopted to calculate the models' outer boundary conditions, and/or by reducing the adopted value of the mixing length parameter with respect to the solar-calibration.}
{A definitive assessment of whether 
the \teff\ scale of metal-poor stellar red giant branch models is consistent with observations requires 
a more robust determination of the fundamental parameters of HD~122563 and also a larger sample of calibrators. From the theoretical side, it is crucial to minimise the current uncertainties in the treatment (boundary conditions, temperature gradient) of the 
outer layers of stellar models with 
convective envelopes.}

\keywords{stars: individual: HD\,122563 -- stars: late-type -- stars: low mass -- stars: fundamental parameters 
}

\titlerunning{The effective temperature scale of metal-poor RGB stellar models}
\authorrunning{O.L.~Creevey et al.}
   \maketitle
%
%________________________________________________________________

\section{Introduction}\label{intro}

Red Giant Branch (RGB) stars are crucial objects for addressing {\sl a plethora} of astrophysics questions.  For example, 
distances to stellar systems can be determined from the brightness of the RGB tip \citep[see, e.g.,][and references therein]{serenelli:17}, or metallicities  of complex stellar systems can be inferred from the RGB location and slope 
\citep[see, e.g.,][]{zoccali:03}. 
Furthermore, RGBs are used for constraining the properties of \lq{exotic}\rq\ particles \citep[see, e.g.,][]{castellani:93} and in general non-standard Physics \citep[see, e.g.,][]{zurab:06, extra:00}. 
In recent years, the huge amount of data on field RGB stars collected by asteroseismic surveys -- in particular by the $Kepler$ mission \citep{keplermission,garcia11,deridder16,mathur16,miglio16,mathur22,vrard22,kuszlewicz23} shows that these objects can be employed as a major tool for Galactic archaeology studies \citep[e.g.][]{victor18,chaplin20,miglio21}.
The theoretical modelling of RGB stars plays therefore a wide-ranging role, involving various fields of both stellar, Galactic and extra-galactic astronomy. 

The properties of evolutionary RGB models critically depend on the input physics and assumptions adopted in the calculations \citep[we refer to the review by][for a detailed discussion on this issue]{scw:02}. Particularly important is the accuracy of their effective temperatures, because it directly impacts our ability to constrain photometrically the metallicity distribution of both simple and complex stellar systems, as well as to constrain the properties of the targets in asteroseismic investigations \citep[see, e.g.,][and references therein]{creevey:12,creevey:19}. 

As it is well known \citep[e.g.][]{scw:02,cs:13}, for a given chemical composition, the \teff\ of RGB models depends on the low-temperature radiative opacity, equation of state, photospheric boundary conditions, and the treatment of the convective efficiency in the outer super-adiabatic layers, which fixes the value of the temperature gradient.
In particular, these last two items are subject to uncertainties that are difficult to quantify a priori.

In almost all stellar evolution codes, the superadiabatic convective temperature gradient is calculated using the simple 
formalism (in the stellar envelopes' regime) provided by the mixing length theory \citep[MLT,][]{bv:58}. 
In the MLT, all relevant physical quantities are evaluated locally, 
and the calculation of the local temperature gradient depends on the value of the mixing length $\Lambda = \alpha_{MLT}\times{H_P}$., where $H_P$ is the local pressure scale height, and $\alpha_{MLT}$ is a free parameter commonly assumed to be a constant value within the convective regions and along all evolutionary phases of stars of any initial chemical composition.
The value of $\alpha_{MLT}$ 
is usually calibrated by reproducing the radius of the Sun at the solar age with a theoretical solar standard model \citep[SSM -- see, e.g.][for a discussion on the SSM calibration]{bastiac:18}\footnote{Additional free parameters are embedded in the MLT formalism, but they are generally fixed a priori, giving origin to different \lq{flavours}\rq\ of the MLT \citep[see, e.g.,][and references therein]{sc:08}. Different MLT flavours provide essentially the same evolutionary tracks once the parameter $\alpha_{MLT}$ is accordingly calibrated on the Sun via the SSM.}. 
There is however no guarantee that a single value of $\alpha_{MLT}$  (in this case the solar value) is also appropriate for stars in different evolutionary stages and/or with different initial chemical compositions, where the mass thickness, and pressure/temperature stratifications of the superadiabatic layers are different compared to the Sun.  

Also, the choice of the outer boundary conditions (BCs) of the models -- pressure and temperature, T, at a given optical depth $\tau$ -- is subject to uncertainties. 
In stellar model calculations either the integration of a $T(\tau)$ relation for the atmospheric layers 
down to a chosen value of $\tau$ is performed, or pressures and temperatures at a given $\tau$ are taken from detailed independent model atmosphere calculations.   Additionally, concerning the $T(\tau)$ relations, there are various choices in the literature.  
Different choices for the BCs imply different solar calibrations of $\alpha_{MLT}$ and lead to different \teff\ scales for the RGB models \citep[see, e.g.,][]{scw:02}.
It is in fact the combination of the choices for the BCs and the calibration of $\alpha_{MLT}$ that affect in a major way the \teff\ of RGB models.

RGB stars, especially in the metal-poor regime (far away from the solar metallicity 
of the $\alpha_{MLT}$ calibration) are therefore very important to test stellar models, especially in relation to the choices of BCs and $\alpha_{MLT}$.
Indeed, there have been already several works devoted to testing RGB models with 
solar calibrated $\alpha_{MLT}$ on field and Galactic globular cluster RGB stars 
\citep[see, e.g.][]{sc:96,scw:02,cs:13,tayar,scsp} with somehow contradicting results.

One particularly interesting target that can be used to test RGB models 
is the star HD~122563: it is one of the brightest, nearby, metal-poor RGB stars, and it has been investigated with several independent, but complementary,  methodologies ranging from spectroscopy, to interferometry, photometry and asteroseismology \citep[see, e.g.,][and references therein]{creevey:19}. 
Due to the availability of a complete and robust observational dataset, this star is 
considered to be an important benchmark for testing stellar physics, such as non-local thermodynamics and 3D effects in model atmosphere computations \citep{amarsi:16}, as well as being a reference calibrator for determination of stellar parameters from large surveys aimed at Galactic archeology \citep[see, e.g.,][and references therein]{gilmore,jofre,gbs23}.

A first careful test of stellar evolution models against the observational data for HD~122563 was performed by \cite{creevey:12,creevey:19}. The authors found that there was a significant discrepancy between the observed and predicted position of this star in the H-R diagram.
Since then observations have improved -- mainly related to the new Gaia data release \citep{gaiamission,gaiadr3summary}, and we take this opportunity to gain a fresh perspective on the comparison between low-metallicity RGB models and this star. 

The paper is laid out as follows: in Section~\ref{obs} we present the observational properties of HD~122563 relevant to this work, and Sect.~\ref{theo} describes the 
reference stellar models used in our analysis. Section~\ref{comp} focuses on the comparisons between theory and observations, including a number of 
numerical experiments and 
evolutionary tracks from other widely used model libraries.
Conclusions and final remarks are presented in Sect.~\ref{end}.

\section{HD~122563 stellar parameters}\label{obs}

HD~122563 (Gaia DR3 3723554268436602240, $G=5.9,V=6.2$) is a star which has been extensively studied with many independent observational techniques, that have provided robust and accurate empirical estimates for some of its fundamental properties, although not all of them, as we will see shortly. 
Here, we simply select what we think are the most accurate and reliable data for an updated comparison with the theoretical framework.

\subsection{Effective temperature}

As extensively discussed in \citet[][hereinafter C19]{creevey:19}, there are many independent measurements of the \teff\ of this star, and almost all of them are in extremely good agreement. Using independent  interferometric measurements,  \citet[][hereinafter C12]{creevey:12} determined $\teff=4598\pm 41$~K, while \citet[][hereinafter K18]{k:18} derived $4636 \pm 36$~K. By employing the infra-red flux method \cite{casagrande:14} estimated
$\teff=4600\pm 47$~K, while a re-analysis of several metal-poor stars including HD\,122563 by \citet[][hereinafter K20]{karovicova20} has provided $\teff = 4635 \pm 34$~ K. 

The interferometric determinations depend on the observed angular diameter and the adopted bolometric flux, this latter being  almost identical in all analyses, and further confirmed by \citet[][hereinafter S23]{gbs23} who independently derived \fbol\ = 13.23 \pmm\ 0.24 erg s$^{-1}$ cm$^{-2}$ using a homogenous compilation of many photometric datasets including data derived using the published Gaia DR3 XP spectra \citep{gaiadr3-montegriffo}.  In that paper, \citeauthor{gbs23} used the K20 interferometric value and derived \teff\ = 4642 \pmm\ 35 K.
We summarise these references in Table~\ref{tab:tefflit}.

The agreement between all the independent measurements is extremely good, and the associated errors are extremely small. To encompass the whole range of measured \teff\ values, in this work we adopt the two extreme estimates by C12 and S23.

\begin{table}[h]
    \centering
       \caption{Sources of \teff\ measurements}
    \begin{tabular}{llllll}
    \hline\hline
    Reference & $\theta$ & \fbol\ & \teff \\
    & [mas] & [erg s$^{-1}$ cm$^{-2}$ 10$^{-8}$] & [K] \\
    \hline
        C12 & 0.940 \pmm\ 0.011 & 13.16 \pmm\ 0.36 & 4598 \pmm\ 41 \\
        K18 & 0.926 \pmm\ 0.011 & 13.20 \pmm\ 0.29 & 4636 \pmm\ 36 \\
        K20 & 0.925 \pmm\ 0.011 & 13.14 \pmm\ 0.22 & 4635 \pmm\ 34 \\
        S23 & using K20 & 13.23 \pmm\ 0.24 & 4642 \pmm\ 35 \\
         & \\
        C14 & & & 4600 \pmm\ 47 \\ 
\hline
\end{tabular}
    \label{tab:tefflit}
\end{table}

\subsection{Bolometric luminosity}\label{sec:lum}

To compare theory and observations in the traditional Hertzsprung-Russell (H-R) diagram, we also need an estimate of the bolometric luminosity of our target. Two different approaches have been used:  
parallax measurements and asteroseismology \citep{creevey:19}.

\subsubsection{Luminosity from parallax measurements}
The Gaia mission \citep{gaiamission} and its data analysis provide astrometric measurements for each observed source in the sky.   One of the parameters that is solved in the astrometric solution is the parallactic motion of the star due to its projected displacement on the sky with reference to the background stars.  
Gaia Data Release 2 \citep{gaiadr2summary} provided the first measurements of parallax based on 22 months of data and derived a value of 3.44 \pmm\ 0.066 mas for HD\,122563.  Part of the analysis in C19 is based on this value.   However Gaia continuously scans the sky and with more measurements over a longer baseline the parallax solution becomes more precise.   Gaia Data Release 3 \citep{gaiadr3summary}, released in June 2022 is based on 34 months of Gaia observations and  the new parallax is 3.099 \pmm\ 0.033 mas, smaller than the previously published one, placing the star further away, hence more intrinsically luminous.
Given this significant change in the parallax value, its luminosity needs to be revisited.  To do this we use the \fbol\ values from C12 and S23, along with the latest parallax and the standard equation
\begin{equation}
    \mathcal{L} = 4 \pi d^2 \fbol,
\end{equation}
where $d$ is the distance to the star.   The parallax is the inverse of the distance, and because the relative error on the parallax is small, we do not use a prior to infer the distance \citep[see e.g.][for a discussion on this topic]{luri18}.   We perform a very simple bootstrap method to derive the luminosity by perturbing the observational data for $N=10\,000$ simulations, and the resulting distributions give  
$L_{\rm C12} =  428^{+15}_{-14}$ \lsol\ and $L_{\rm S23} = 431^{+12}_{-12}$ \lsol\, respectively.

We note that the Hipparcos mission was the first to record a parallax for HD\,122563, equal to 4.22 \pmm\ 0.36 mas, and this value was used in C12, when a first detailed comparison with stellar models was made. We also derive the radius and  surface gravity of the star, where for the latter we assume that the star's mass is centred on 0.80 \msol\ with an error of 0.05 \msol, consistent with the star being a metal-poor, halo giant with an age on the order of $\sim12.5$~Gyr..

\subsubsection{Luminosity from asteroseismology}

Asteroseismic analyses were performed on a data set of radial velocity measurements for HD\,122563 observed with the SONG telescope \citep{songrefa, songrefb}, and in C19 a value of the asteroseismic quantity \numax\ (= $3.07 \pm 0.05$~$\mu$Hz) was measured.   This quantity is proportional to the surface gravity $g$ of the star, and if we assume a mass, then we can derive its radius.  As the angular diameter is known, we can then derive an {\it asteroseismic distance} independent of a distance measurement.  

We follow here this same approach by using both the C12 and S23
\teff\ values along with the C19 \numax. 
We calculate $g$ as :
    \begin{equation}
\frac{\numax}{\nu_{\rm max\odot}} = f_{\nu_{\rm max}}\frac{g}{g_\odot} 
\sqrt{ \frac{T_{\rm eff\odot}}{\teff} } 
\label{eqn:numax}
\end{equation}
where $f_{\nu_{\rm max}}$ = 1.0 and $\nu_{\rm max,\odot}$ = 3~050 $\mu$Hz \citep{kb95}.  
In this work we adopt $T_{\rm eff,\odot} = 5772$ K and $\log g_{\odot}$ = 4.438 dex from the IAU convention\footnote{\url{https://www.iau.org/static/resolutions/IAU2015_English.pdf} Resolution B3} 
\citep{iau15b3}.

The equation for surface gravity is 
\begin{equation}
    g = \frac{GM}{R^2},
\end{equation}
which can be rewritten as
\begin{equation}
\frac{g}{g_\odot} = \frac{M}{M_\odot} \left(\frac{R_\odot}{R}\right)^2.
\label{eqn:grav2}
\end{equation}
Here \rsol\ = $6.957 \times 10^{10}$ cm, while for $\frac{M}{M_\odot}$ we again assume a distribution centred on $0.80 \pm 0.05$ \msol.
In this way $R$ can then be derived, and finally the distance. 

To estimate the stellar properties and uncertainties, we performed simulations and used their distributions to determine the asteroseismic luminosities and distances.  These are given in the middle panel of 
Table~\ref{tab:obs}; 
the star appears to be less luminous and closer compared to using the parallax measurement. 

It is interesting to investigate also the impact of 
changing the assumption on the value of the stellar mass, and the value of \numax. 
If we assume a mass centred on 0.85 \msol\ instead of 0.80 \msol, the effect is to increase the radius and thus the derived distance and luminosity at fixed \numax.  
A shift by +/--0.05 \msol\ increases/decreases the luminosity by $\sim$6\% and the distance by $\sim$3\%.  
Increasing/decreasing the \numax\ value by 1$\sigma$ will decrease/increase the luminosity by 1.5\% and the distance by $<1$\% at fixed mass.

There is additionally an empirical uncertainty on the value of $f_{\numax}$ in Eq.~\ref{eqn:numax}. It is under debate whether the classical scaling law for $\nu_{max}$ should be modified to account for the chemical and structural differences in the outer layers of stellar targets compared to sun-like stars \citep[we refer to][and references therein]{viani:17}.  These corrections, however, change the derived luminosity and distance by smaller amounts compared to the effect of the offsets in mass 
and \numax\ described above.

We therefore decided to be conservative and here we presently adopt the determination of the HD~122563 luminosity obtained via the classical (not accounting for second-order corrections) scaling law. 
Table~\ref{tab:obs} summarizes the various determinations of the fundamental 
parameters of this star.

\begin{table*}
\caption{Fundamental parameters of HD~122563 (Gaia DR3 3723554268436602240).  The top panel shows the existing literature data while the middle and lower panels show the results from this work.}
\label{tab:obs}
\centering
\begin{tabular}{l c c c c}
\hline\hline
Property & \multicolumn{2}{c}{Creevey et al. (2012)} &  \multicolumn{2}{c}{K20+S23} \\
\hline
\fbol\ [erg~s$^{-1}$~cm$^{-2}$]& \multicolumn{2}{c}{13.16 \pmm 0.36e-8 } & \multicolumn{2}{c}{13.23 \pmm\ 0.24e-8} \\
$\theta$ [mas] & \multicolumn{2}{c}{0.940 \pmm\ 0.011} & \multicolumn{2}{c}{0.925 \pmm\ 0.011} \\
\teff~[K] & \multicolumn{2}{c}{$4598\pm41$} &  \multicolumn{2}{c}{$4642\pm35$} \\
$\pi$~[mas] & \multicolumn{2}{c}{3.099$\pm$0.033}\\

\hline
              & Seismology & Gaia DR3 parallax & Seismology & Gaia DR3 parallax \\
\hline
   $L$ [L$_\odot$] & $359^{+26}_{-25}$ &  $428^{+15}_{-14}$  &  $370^{+26}_{-26}$ &  $431^{+12}_{-12}$\\
   $R$ [R$_\odot$] & $29.8\pm1.0$ & $32.6 \pm 0.5$ & $29.8\pm1.0$ & $32.1 \pm 0.5$\\
  $\log{g}$~[dex] &   $1.392\pm 0.007$ & $1.32\pm0.03$ & $1.394\pm 0.007$   & $1.33\pm 0.03$\\
  $d$ [pc] & $295\pm 10$ & 322.7$\pm$3.5 & 300$\pm$10 & -- \\
  \hline
  \hline
$\feh_{spec}$~[dex] & $-2.43\pm0.15$ & \multicolumn{3}{c}{}\\
$\feh_{phot}$~[dex] & $-2.30\pm0.15$ & \multicolumn{3}{c}{}\\

$\alfe$~[dex] & $+0.4\pm0.1$ & \multicolumn{3}{c}{}\\
\hline
\end{tabular}

\end{table*}

\subsection{Iron abundance and heavy elements distribution}\label{feh}

The exact chemical composition of a stellar target is a critical ingredient when using this object as a benchmark for stellar models; this is because the effective temperature scale of stellar models -- and this is particularly true in case of cool giant stars -- critically depends on the iron content as well as $\alpha-$element abundances.   Iron is an important opacity source in the low-temperature regime (hence for the cool outer layers of giants), while $\alpha-$elements, and in particular Mg, Si and O (in order of importance), due to their low energy ionization levels are fundamental electron donors and, hence have a huge impact on the $H^-$ ion opacity that is the major opacity contributor in the cool envelope of RGB stars \citep[][]{cs:13}.

Concerning the iron abundance \feh \, of HD\,122563 there is a large spread of values in the literature. 
The estimates range from $\feh=-2.92$ as provided by \cite{afsar}, to $-$2.75 from \cite{karovicova20}, $-2.7$ from \cite{collet:18}, 
--2.64 from \cite{jofre14}, $\sim-2.5$ as given by \cite{praka:17}, and $\feh=-2.43$ \citep{amarsi:16}.
While part of this large spread of [Fe/H] values is due to the chosen reference solar composition adopted in the  analyses, most of the variations are associated with the use of different methodologies and/or observational techniques. 

A very accurate analysis of the chemical abundances for HD~122563 is the one by \cite{amarsi:16}, who made use of 3D non-LTE radiative transfer calculations based on the (3D) hydrodynamic STAGGER model atmospheres. 
They determined both LTE and non-LTE abundances from  both FeI and FeII lines, and in this paper we adopt the stable abundance from the FeII lines, equal to $\feh=-2.43$, as listed above.  FeII lines are less affected by non-LTE effects and this estimate is therefore more reliable.  
However, we must keep in mind a conservative uncertainty of about 0.15~dex on this value.

Just as for [Fe/H], a large spread also exists concerning the 
$\alpha$-element abundances.  Significant differences in the literature stem from the adopted model atmospheres (1D versus 3D) and selected spectroscopic lines \citep[we refer to][for a discussion on this point]{praka:17, collet:18}.
\cite{collet:18} obtain [O/Fe] varying from +0.08 
from molecular lines when compared to FeII, to +0.93 from the 630nm line when compared to FeI.
They also derive [N/FeI] = +0.68 and [N/FeII] = +0.29.
\cite{praka:17} derive [O/Fe] = +0.07 to +0.37 using OH UV, IR, or the forbidden [OI] line.
\cite{jofre15} derive [Mg/Fe]$=+0.29$, [Si/Fe]$=+0.32$ and  [Ca/Fe] = +0.21, and \cite{afsar} derive [Mg/Fe]$=+0.46$, [Si/Fe]$=+0.45$ and [Ca/Fe] = +0.38.
More recently in Gaia DR3, \cite{creevey23} derive [Ca/Fe] = +0.33, see \cite{recioblanco23} for details.  
In general, we find that the \alfe\ corresponds to  approximately +0.4 dex.

%--------------------------------------------------->
   \begin{figure}
     \centering
     \includegraphics[width=9.0cm]{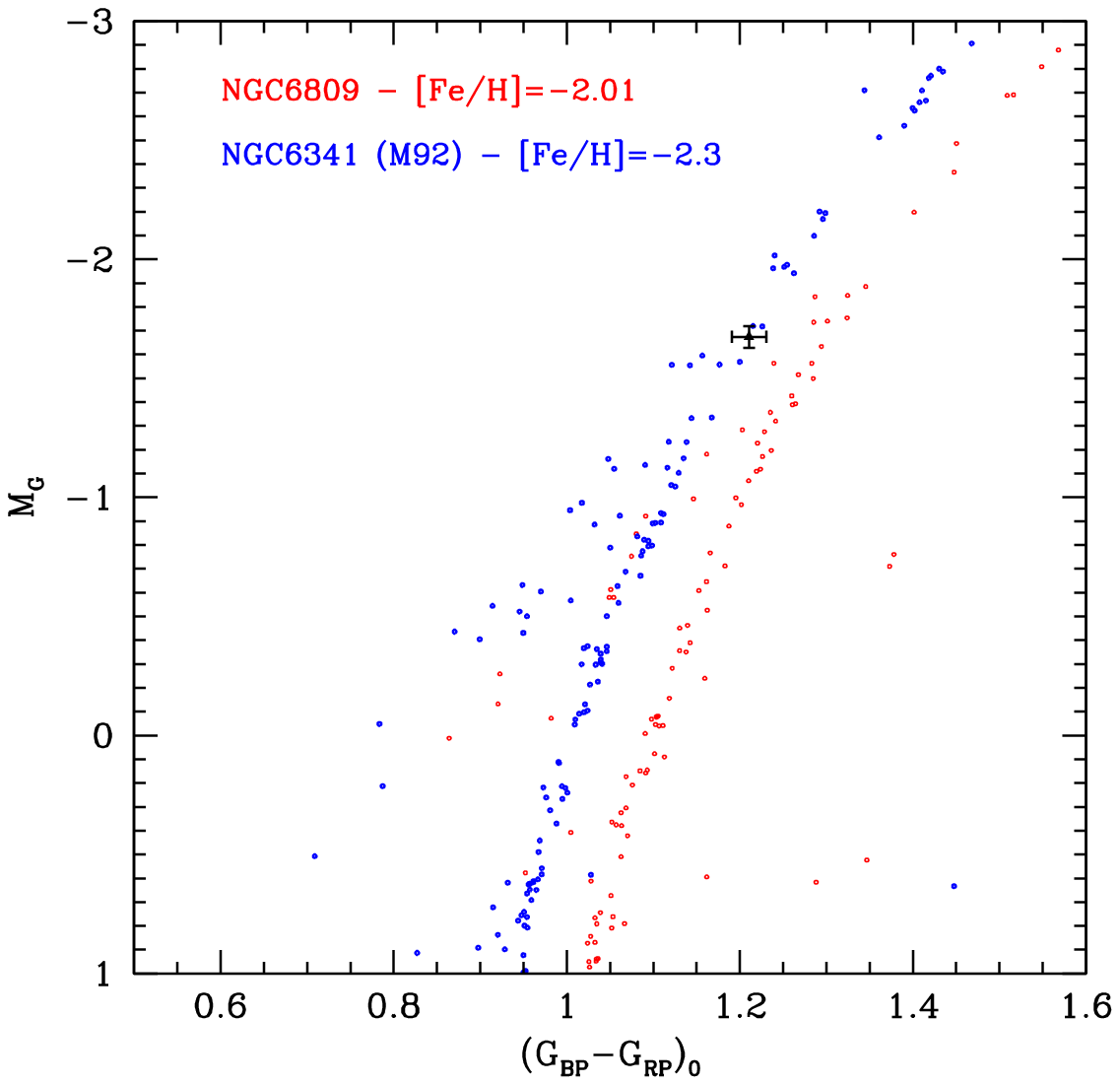}
     \vskip -0.5truecm
      \caption{Comparison in the $M_{\rm G} - (G_{\rm BP}-G_{\rm RP})_0$ CMD between the GCs M92 and NGC~6809 and the location of the star HD~122563 (see Sect.~\ref{obs} for more details).}
         \label{fig:gcs}
   \end{figure}
%---------------------------------------------------|

\subsection{Photometric metallicity from a comparison to globular clusters}\label{sec:absmag}

As an independent check of our adopted chemical 
composition, we have  compared the position of this star in the (distance and reddening corrected) Gaia Colour-Magnitude diagram (CMD) with the RGBs of selected Galactic globular clusters (GCs) with high-resolution spectroscopic determinations of their chemical composition. We employed NGC~6341 (M92), with $\feh=-2.30$ \citep[see, e.g.][and references therein]{lee:23}, and NGC~6809 with an iron abundance equal to $-2.01$ \citep{rain:19}.  Both clusters have an $\alfe\approx+0.40$, similar to the $\alpha-$enhancement of HD~122563.

To derive \mg\ of HD\,122563 
we use the standard observational approach 
\begin{equation}
    \mg = G + 5 - 5\log_{10}(\varpi) - A_G,
\end{equation}
where $G$ is the observed $G$ magnitude from Gaia, $\varpi$ is the parallax in mas, and $A_G$ is the extinction in $G$ band.    We employed the published $A_V = 0.01 \pm 0.01$ mag for the extinction and converted it to $A_G$ using the equations provided by \cite{danielski18} with the updated coefficients for DR3 given in the Gaia DR3 software webpages\footnote{\url{https://www.cosmos.esa.int/web/gaia/edr3-extinction-law}}.  This equation requires as  input the extinction defined at $\lambda = 550$ nm, called $A_0$, and for this we used the value of $A_V$.  We employed the coefficients based on \teff\ and \feh\, 
using the adopted values described in the previous sections. 

To derive the dereddened colour \bprpzero = \bprp - \ebpminrp, we applied the same method to estimate the extinction in the $G_{\rm BP}$ and $G_{\rm RP}$ bands --  $A_{\rm BP}$ and $A_{\rm RP}$ -- and then subtracted the colour excess \ebpminrp\ from the observed $(G_{\rm BP} - G_{\rm RP})$.   
These quantities are summarised in Table~\ref{tab:hd122563_cmd}.

\begin{table}[]
    \centering
        \caption{Absolute magnitude and intrinsic colour for HD\,122563.}    \begin{tabular}{cccc}
    \hline \hline 
    \teff\ & $A_V$ & $M_G$ & \bprpzero\ \\
    \hline
    4598 \pmm\ 41  & 0.01 & --1.680 \pmm\ 0.024  &  1.21073 \pmm\ 0.00003 \\
    4642 \pmm\ 35 & 0.01 &  --1.679 \pmm\ 0.024  &  1.21070 \pmm\ 0.00003   \\
    \hline\hline
    \end{tabular}
    \label{tab:hd122563_cmd}
\end{table}

As for the GCs, 
we used the Gaia data to select members and derive intrinsic magnitudes and colours for M92 and NGC\,6809. 
We first selected all stars within a search radius around the centre of both clusters and calculated a distance in position $d_{\rm pos}$ and proper motion $d_{\rm pm}$ from the median values of the sample of stars, and then filtered the members based on a limit in both $d_{\rm pos}$ and $d_{\rm pm}$.
Once filtered on position and proper motion, we removed noisy sources or duplicated sources by filtering on\\
{\tt ipd\_frac\_multi\_peak}\\ {\tt ruwe}, 
{\tt phot\_bp\_mean\_flux\_over\_error} \\
{\tt phot\_rp\_mean\_flux\_over\_error} \\
{\tt phot\_g\_mean\_flux\_over\_error}, {\tt phot\_bp\_n\_obs} \\
and 
{\tt phot\_bp\_n\_blended\_transits}.  

The actual limits imposed were based on inspection of the observed CMD to ensure that we retained sufficient members of the cluster in critical areas of their evolution stage, and removed as many outliers as possible.  
The details of the values used are given in Table~\ref{tab:clusterfiltering}.
We note that to derive the distance we inspected the cluster members along the horizontal branch, by properly accounting for reddening at those \teff,  and then adopted the distance needed to match the Zero Age Horizontal Branch predicted by models with the clusters' [Fe/H]  from the $\alpha$-enhanced BaSTI library \citep[][]{pietrinferni:21}. This distance\footnote{We note that the estimated distance for the two GCs are in fair agreement with the estimates provided by {Vandeberg (2023)} by using the same approach but his own ZAHB models and different photometric data: $\sim8520$~pc and $\sim5200$~pc for M92 and NGC~6809, respectively.} was then used for the calculation of \mg\ of the member stars. It is worth noting that due to the vertical shape of the RGB -- especially in the very metal-poor regime -- a reasonable uncertainty of $\sim0.10-0.15$~mag in the GC distance moduli would not significantly affect the results of our comparison with HD~122563.

We have corrected the clusters' photometry for extinction 
following the same procedure as for our star. For 
$A_0$ we used the values of $A_V$ calculated from 
$E(B-V)$ given by \citet{harris96}, using $R_V=3.1$. 
The extinctions were calculated for a reference 
\teff\ = 4600~K.   As Table~\ref{tab:clusterfiltering} implies, varying the \teff\ by $\pm 100$ K has a negligible impact on the adopted extinction.

\begin{table}[]
    \centering
        \caption{Properties and filtering of globular cluster members.  The \ebv\ is taken directly from \cite{harris96}.}
    \begin{tabular}{rccccc}
    \hline\hline
    Parameter & M92 & NGC\,6809 \\
    \hline
    \feh &  --2.31 & --2.01\\
    RA [deg] &  259.28125 &  294.9996\\
    DEC [deg] & 43.13595 & -- 30.9625\\
    pmRA [deg] & -4.8831 & --2.7813\\
    pmDEC [deg] & -0.67986 & --7.5716\\
    $d_{\rm pos}$ [arcmin] [$<$] & 4.8 & 6.6\\
    $d_{\rm pm}$ [arcmin] [$<$] & 6.0 & 3.6\\
    {\tt ipd\_frac\_multi\_peak} [$<$] & 3 & 2 \\
    {\tt ruwe} [$<$] & 1.4& 1.4  \\ 
{\tt phot\_bp\_mean\_flux\_over\_error} [$>$] & 5 & 10 \\
{\tt phot\_rp\_mean\_flux\_over\_error} [$>$]& 4 & 10 \\
{\tt phot\_g\_mean\_flux\_over\_error}[$>$] & 100 & 150\\
{\tt phot\_bp\_n\_obs} [$>$]& 5 & 5\\
{\tt phot\_bp\_n\_blended\_transits}[$<$] & 10 & 3\\
{\tt phot\_rp\_n\_blended\_transits}[$<$] & -- & 3\\
$N_{\rm stars}$         & 3597 & 3330 \\
$E(B - V)$      [mag]   & 0.02 & 0.08 \\
\ebpminrp (@4600 K) [mag]& 0.0243 & 0.097  \\
$A_G$ (@4600 K) [mag] & 0.0500 & 0.1990\\
\ebpminrp (@8000 K) [mag]& 0.0295  & 0.1172  \\
$A_G$ (@8000 K) [mag]& 0.0594 & 0.2361\\
$d$ [pc] & {8700} & {5900}\\
\hline \hline 
    \end{tabular}
    \label{tab:clusterfiltering}
\end{table}

The data for both globular clusters and HD\,122563 are shown in Fig.~\ref{fig:gcs}. The position of HD~122563 in the Gaia CMD is consistent with the iron abundance estimated for M92, 
which overlaps with our adopted [Fe/H] by \citet{amarsi:16} 
within the associated error bar.

{We note that if the measured extinction towards M92 turned out to be underestimated, HD122563 then would  appear more metal-rich.   However, if the assumptions on the extinction towards NGC6809 were incorrect,  the impact would be to simply increase or decrease the uncertainty on the photometric metallicity.}

%--------------------------------------------------->
   \begin{figure}
     \centering
     \includegraphics[width=9.5cm]{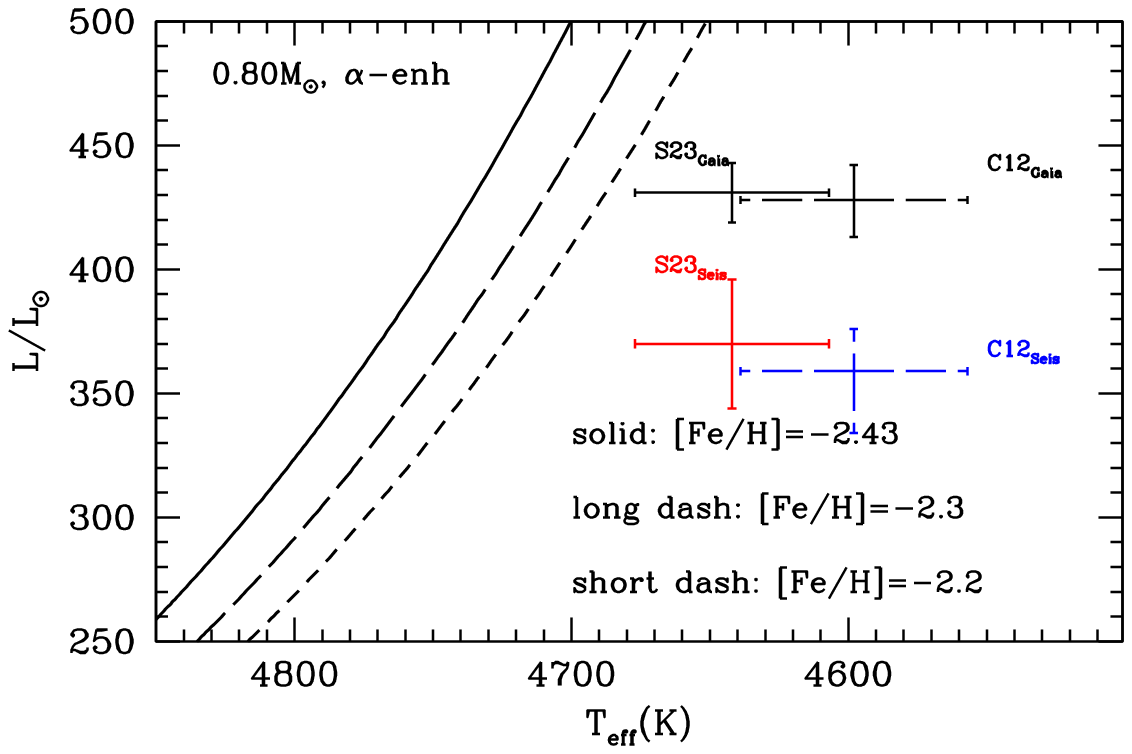}
     \vskip -2.5truecm
      \caption{H-R diagram of HD~122563. The different points with error bars correspond to the empirical measurements based on the two interferometric \teff\ estimates and the different methods for fixing the distance to the target (see Sect.~\ref{sec:lum} for more details). The various lines are the RGB evolutionary tracks of $0.8~M_\odot$ $\alpha-$enhanced stellar models with 
      the labelled values of $\feh$.}
         \label{fig:hr1}
   \end{figure}
%---------------------------------------------------|

\section{Stellar evolutionary tracks}\label{theo}

The baseline stellar models adopted in this analysis are an updated version of the BaSTI stellar model library initially presented in \cite{basti} and used by C19. The solar-scaled version of the updated models has been published in \cite{bastiac:18}, and represents a significant improvement compared to the previous release of the database\footnote{\bf Unlike previous BaSTI models, the new ones  
include also atomic diffusion. However, 
atomic diffusion does not have any significant impact on the \teff\ of RGB models, because the $1^{st}$ dredge-up basically restores the initial metallicity in the RGB envelopes \citep[see, e.g.,][for a more detailed discussion on this topic]{cs:13}.}. The changes in the input physics that affect the \teff\ scale of the RGB models are the following:

\begin{itemize}

\item an updated solar heavy elements distribution  provided by \cite{caffau}, which gives $(Z/X)_\odot$ = 0.0209 and $Z_\odot=0.0153$.  
Here $X$, $Z$ are, as customary, the mass fractions of hydrogen and of all elements heavier than helium (metals), respectively. The previous BaSTI models employed the \citet{gn93} solar metal 
mixture, which provided a larger value of $Z_\odot$.

\item{outer boundary conditions (BCs) obtained by integrating the atmospheric layers using the $T(\tau)$ relation by \cite{vernazza:81}, while the previous BaSTI models were based on the $T(\tau)$ by \citet[][herinafter K66]{ks}. As discussed in \cite{bastiac:18}, the alternative use of these atmospheric temperature stratifications (after the calibration of the mixing length $\alpha_{MLT}$), 
implies a difference in the \teff\ scale of metal-poor RGB models by about 60~K, the models based on the \cite{vernazza:81} $T(\tau)$ relationship being cooler;}

\item{the efficiency of the superadiabatic convection has been fixed by using the same MLT formalism used in the earlier BaSTI release, but the value of $\alpha_{MLT}$ has been recalibrated by computing a standard solar model (SSM) \citep[see the discussion in][]{bastiac:18}, to take into account the changes in the solar mixture and outer boundary conditions.}

\end{itemize}

Given that we are dealing with a halo, metal-poor $\alpha$-enhanced star, 
we employ here the new $\alpha-$enhanced BaSTI models \citep{pietrinferni:21}, which rely on the same physics inputs as the scaled-solar ones, but employ a heavy element distribution with all $\alpha-$elements homogeneously enhanced by +0.4~dex with respect to Fe, compared to 
the solar-scaled distribution. 

In addition to the baseline stellar models available in the BaSTI library, we have used in our analysis additional dedicated sets of calculations, as discussed below.

\section{Comparison between theory and observations}\label{comp}

Figure~\ref{fig:hr1} shows a comparison between the H-R diagram of our target star and our 
 $0.80M_\odot$  baseline RGB evolutionary tracks for $\feh=-2.43$  (solid line, the same iron abundance adopted from the spectroscopic analysis by \citealt{amarsi:16}), $\feh=-2.3$ (long dashed line) and $\feh=-2.2$ (short dashed line). 
At the luminosity derived from the Gaia distance, the $\feh=-2.43$ track is hotter by about $\sim90$~K and 130~K than the $T_{\rm eff}$ determinations by S23 and C12, respectively. This means that -- thanks mainly to the new Gaia DR3 parallax -- the disagreement between theory and observations has been roughly halved in comparison with the analysis by \cite{creevey:19}.

When considering the models for $\feh=-2.3$ -- the nominal [Fe/H] of M92 and roughly the upper limit to the adopted spectroscopic [Fe/H] -- 
the discrepancy is reduced to about 70~K and 106~K compared to the S23 and C12 $T_{\rm eff}$ estimates, respectively; i.e. at this (photometric-) metallicity, 
the position of HD~122563 in the H-R differs from the models at the level of about $2\sigma$.

Given that in the explored metallicity regime, a change by 0.1~dex in the iron abundance changes the \teff\ scale of the RGB tracks by about $\sim25$~K, our baseline models would agree with 
observations at the $1\sigma$ level by increasing $\feh$ by an additional 0.13 dex above [Fe/H]=$-$2.3 when adopting the S23 $T_{\rm eff}$, or by  0.3~dex in the case of the C12 $T_{\rm eff}$.

\begin{figure}
     \centering
     \includegraphics[width=9.5cm]{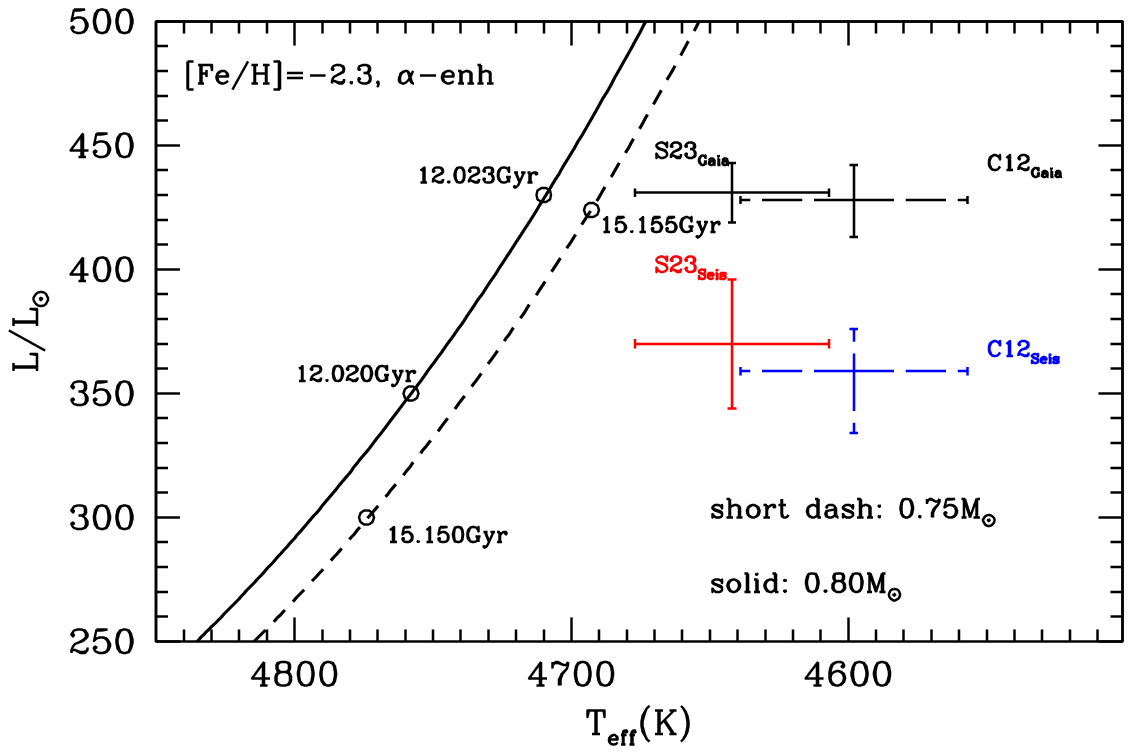}
     \vskip -2.5truecm
      \caption{As Fig.~\ref{fig:hr1}, but in this case the lines correspond to RGB evolutionary tracks for the labelled masses and $\feh=-$2.3. Selected values of the model ages are shown along the tracks.}
         \label{fig:mass}
   \end{figure}

Figure~\ref{fig:mass} shows the same comparison of Fig.~\ref{fig:hr1}, but at fixed $\feh=-2.3$ and for two different masses. It is evident that, for the selected iron abundance, the mass cannot be lower than about $0.78M_\odot$, to have ages consistent with the cosmological age 
of $\sim$13.8~Gyr. Models for $0.78M_\odot$ reduce the discrepancy between theory and observations at $\feh=-2.3$ by only 10~K.

%--------------------------------------------------->
%---------------------------------------------------|

%--------------------------------------------------->
   \begin{figure*}
     \centering
     \includegraphics[width=12.5cm]{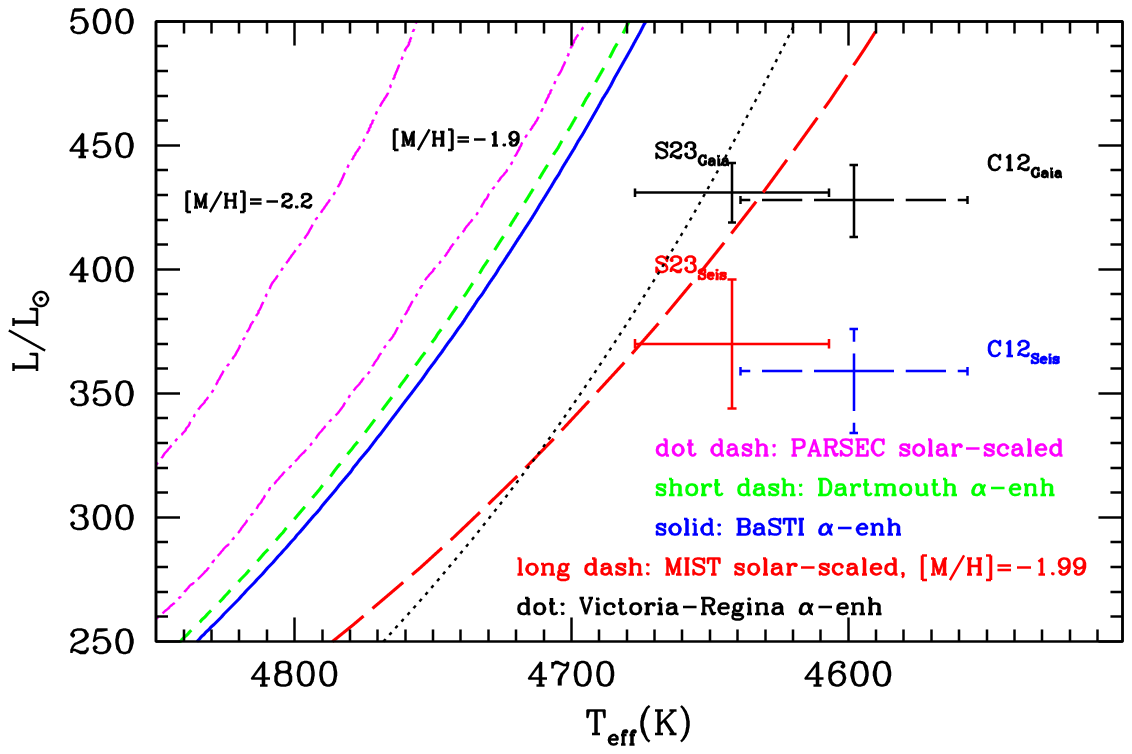}
     \vskip -2.5truecm
      \caption{As Fig.~\ref{fig:hr1}, but in this case, the lines correspond to the RGB $0.8M_\odot$ evolutionary tracks from various stellar model libraries and different assumptions about the chemical composition (see labels and the text for more details). }
         \label{fig:libraries}
   \end{figure*}
%---------------------------------------------------|

\subsection{Comparison with other stellar model libraries}

In addition to our calculations, we have compared the H-R diagram of HD~122563 with recent sets of evolutionary tracks from different groups.
We have selected suitable 0.8$M_{\odot}$ evolutionary tracks from the PARSEC \citep{bressan:2012}, Dartmouth \citep{dotter:07}, MIST \citep{choi:16}, and the Victoria-Regina libraries \citep{don:23}.

While the BaSTI, Dartmouth and Victoria-Regina models are $\alpha-$enhanced, the models from the other two libraries are based on a solar-scaled heavy element distribution, and in the comparison 
we chose tracks with a total metallicity [M/H] 
as close as possible to the metallicity derived from $\feh=-2.3$ and ${\rm [\alpha/Fe]=+0.4}$, and the solar metal distribution adopted in those calculations. 
This is an appropriate approximation, especially for metal-poor compositions, as first shown by \citet{salaris:93}.

In the case of the MIST models we used the online tool to calculate a 0.8$M_{\odot}$ track with the appropriate [M/H] ([M/H]=$-$1.99), while in the case of the PARSEC models we had to use the tracks with [M/H] closest to the appropriate value (namely [M/H]=$-2.2$ and $-1.9$, respectively). 
We compare all of these stellar tracks in Fig.~\ref{fig:libraries}.

The Dartmouth and BaSTI tracks for $\feh=-2.3$ are very similar, the BaSTI track being slightly cooler by just $\sim 5$~K.
The PARSEC tracks (even the one with a metallicity slightly higher than the appropriate value) are hotter than the BaSTI track, whilst the MIST and Victoria-Regina tracks are about 80~K cooler than BaSTI at the expected HD~122563's brightness. Both Victoria and MIST tracks agree 
with the observations within less than $1\sigma$ when the S23 \teff\ is adopted, and the 
MIST track agrees with the data to less than $1\sigma$ also when the C12 \teff\ is used 
together with the $Gaia$ distance.

Given this result, it is worthwhile to try and trace the origin of the difference of the \teff\ scale between the Victoria-Regina and MIST tracks and our BaSTI baseline one: 

\begin{itemize}

\item{The first important difference is the approach used for fixing the BCs of the models. The MIST computations adopt BCs provided by model atmosphere computations \citep[we refer to][for more details]{choi:16}; while the Victoria-Regina models rely on the integration of the solar $T(\tau)$ relation by \cite{hm:74};}

\item{both MIST and Victoria-Regina models employ the solar heavy-element distribution by \cite{asplund:09}, which is different from that adopted in the other selected libraries. PARSEC and BaSTI are based on the \cite{caffau:11} solar metal mixture, while the Dartmouth library adopts the solar heavy element distribution by \cite{gs:98}. The different metal mixture affects directly 
the radiative opacities, and indirectly the mixing length calibration. In fact 
the lower solar metallicity induced by the \cite{asplund:09} metal distribution 
compared to \cite{caffau:11} and \cite{gs:98} requires a smaller value of 
the mixing length to calibrate the solar model and this, in turn, implies a lower 
\teff\ of the RGB models.}
\end{itemize}

In the following, we discuss in more details these two points.

\subsection{The role of outer boundary conditions}

The \teff\ scale of RGB models is significantly dependent on the approach used for fixing the BCs required to solve the set of stellar structure equations \cite[see, e.g.,][and references therein]{scw:02,vdb08,sc15,choi-bc:18}. Although, once the mixing length parameter is re-calibrated by means of an SSM, the impact of using different BCs on the RGB \teff\ scale is reduced.  Significant differences do exist among RGB stellar models based on different choices for the calculation of the atmospheric thermal stratification and 
the BCs. As shown by \cite{scw:02}, and more recently by \cite{choi-bc:18}, the use of different BCs can lead to offsets by $\pm100$~K on the RGB, even if in the models the mixing length $\alpha_{MLT}$  has been properly calibrated to the solar value. 

%--------------------------------------------------->
   \begin{figure}
     \centering
     \includegraphics[width=9.0cm]{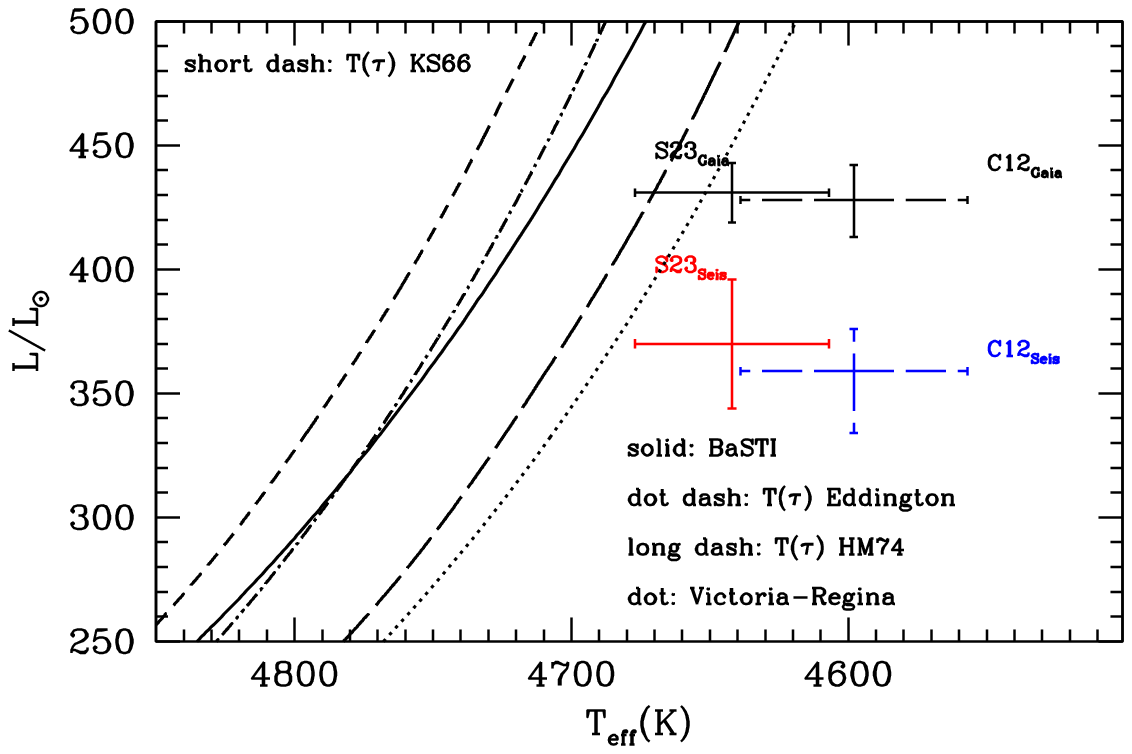}
     \vskip -2.5truecm
      \caption{As Fig.~\ref{fig:ml} but  for various assumptions about the thermal stratification adopted in the atmospheric stellar layers (see labels) in the Basti models. The evolutionary track for the $0.8M_\odot$ model as provided by the Victoria-Regina library is also shown for comparison.}
         \label{fig:bc}
   \end{figure}
%---------------------------------------------------|

We have therefore tested different choices for the BCs,  in the metal-poor regime of our target. To this purpose, we have computed $0.8M_\odot$ evolutionary tracks for  $\feh=-2.3$ by using the same input physics as our baseline BaSTI models, but relying on the integration of different $T({\tau})$ relationships for 
the atmospheric layers. 
We used the KS66 $T(\tau)$, the Eddington (grey) atmosphere, and the  \cite{hm:74} relationship (hereinafter HM74). 
In each case we have fixed the mixing length parameter 
by calibrating a SSM, and we obtained $\alpha_{MLT}= 1.799$ with the KS66 $T(\tau)$, $\alpha_{MLT}= 2.109$ with the HM74 one, and 2.214 for the grey atmosphere.
For comparison, in our baseline models we use $\alpha_{MLT}= 2.006$ 
with the \cite{vernazza:81} $T(\tau)$.

Figure~\ref{fig:bc} shows the results of these calculations, which extend to the very metal-poor regime the analysis performed in \cite{bastiac:18}. The RGBs 
calculated with the grey Eddington $T({\tau})$ are very close to our baseline 
calculations, but with a slightly different slope.
The RGB computed with the KS66 $T(\tau)$ is hotter than the baseline model by about 30~K, a smaller difference (but in the same direction) than the case at higher metallicity, 
while the RGB calculated with the HM74 $T(\tau)$ is cooler than the reference one by $\sim40$~K. This is different than the case at solar metallicity
where RGBs with the HM74 $T(\tau)$ are about 40~K hotter than calculations with the 
 \cite{vernazza:81} $T(\tau)$. 

Figure~\ref{fig:bc} shows that the RGB calculated with the HM74 $T(\tau)$ 
is very close to the RGB of the Victoria-Regina models (differences by 
less than 20~K at the luminosity of our target star). This also implies that 
the impact of the different solar metal mixture is minor, and 
if we use the results of the next section about the variation of the RGB \teff\ induced by a given variation of $\alpha_{MLT}$, we find 
that the remaining temperature difference is due to the different solar calibrated $\alpha_{MLT}$ in the Victoria-Regina models --smaller by 0.1 than our calibration with the HM74 $T(\tau)$--  
induced by the different choice of the solar metal mixture.  Based on these results, we can predict that the use of the solar heavy-element distribution recently suggested by \cite{magg} -- which provides a very similar but slightly larger solar metallicity than that by \cite{caffau:11} -- would not have any significant impact on our analysis.

It seems therefore that one way to reduce/erase the discrepancy between our baseline BaSTI models and the observations of this star is basically to employ a different $T(\tau)$ relation to calculate the BCs. It is however worth recalling that 
\cite{sc15} have shown that RGB models calculated with the \cite{vernazza:81} $T(\tau)$ relation and a solar calibrated mixing length seem adequate to reproduce the effective temperature of RGB stars with $\alfe=0.0$ in the metal-rich regime.

Based on these tests, the interpretation of the results with the MIST model is however less clear. As extensively discussed by \cite{choi-bc:18}, the thermal stratification provided by the model atmospheres used in \cite{choi:16} is well reproduced by the \cite{vernazza:81} $T(\tau)$. But it is then hard to understand the offset and the different 
slope of the RGB compared to the BaSTI models that use the \cite{vernazza:81} $T(\tau)$, given also the small impact of the different solar metallicity, as deduced by the comparison 
with the Victoria-Regina models.

\subsection{The value of $\alpha_{MLT}$}

An alternative possibility to explain the disagreement between the \teff\ of the BaSTI (but also Dartmouth and PARSEC) RGBs and the target star, assuming that their choices to fix the BCs and possibly the solar metal mixture are the \lq{correct\rq} ones, is a variation of 
$\alpha_{MLT}$ compared to the solar calibrated value.   There is abundant literature on this subject, and the reader can refer to the recent review by \cite{jt}.

In a recent work based on APOKASC data, \cite{scsp} found that for stars 
with solar-scaled heavy element composition in 
the [Fe/H] range between $\sim$0.4 and $\sim -$0.6, there is a good agreement between theoretical and observed \teff\ with models consistent with the baseline BaSTI calculations, calculated with solar calibrated $\alpha_{MLT}$. However, the same models appeared to be too hot when compared to stars 
with $\alpha$-enhanced composition in the range $-0.7<\feh<-0.35$~dex. This result is qualitatively consistent with what we find in the comparison of our baseline BaSTI models with HD~122563.

\begin{figure}
     \centering
     \includegraphics[width=9.5cm]{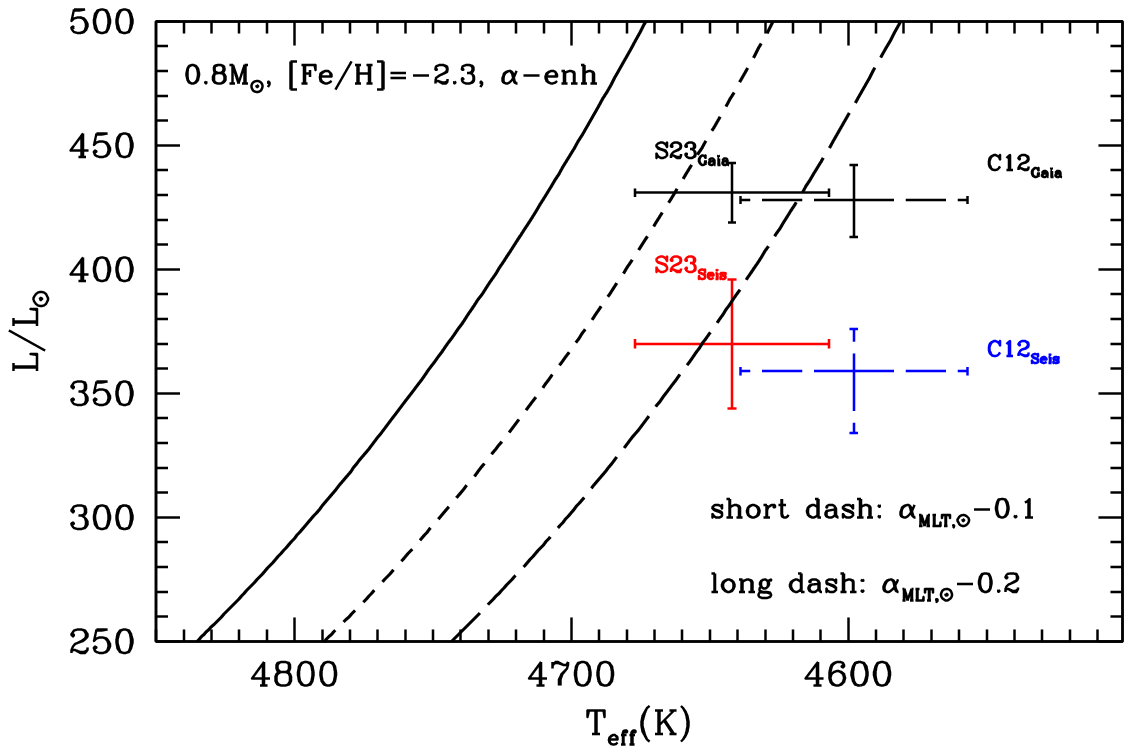}
     \vskip -2.5truecm
      \caption{Same as Fig.~\ref{fig:hr1} but for $0.8M_\odot$ evolutionary computed reducing by various amounts  the calibrated solar  $\alpha_{MLT,\odot}=2.006$ (see labels).}
         \label{fig:ml}
   \end{figure}

To determine what variation of $\alpha_{MLT}$ is required to bring the baseline BaSTI models in better agreement with the observations of HD~122563, 
we have calculated additional 0.8$M_{\odot}$ $\alpha$-enhanced tracks 
with $\feh=-2.3$ and mixing length reduced from the solar value $\alpha_{MLT,\odot}$ by two different amounts, as shown in Fig.~\ref{fig:ml}.  

%--------------------------------------------------->
%---------------------------------------------------|
      
We find that at this metallicity and in the relevant range of luminosities,  
a reduction of $\alpha_{MLT}$ by 0.1 decreases the model \teff\ by $\sim$+45~K.  Hence, a decrease of $\alpha_{MLT}$ by $0.1-0.2$ is 
in principle able to put the BaSTI models in agreement with the observations, for a target metallicity \feh=$-$2.3.

A similar result, i.e. the need for a sub-solar $\alpha_{MLT}$ to better reproduce the \teff\ of metal-poor stars, has been found by \cite{joyce:18a} for a sample of MS and SGB stars by using their own evolutionary code.

\section{Conclusions}\label{end}

From the previous analysis, it appears clear that 
if HD~122563 $\feh \sim -2.3$ (compatible with the metallicity of M92), both Victoria-Regina and MIST models are able -- within the estimated uncertainties on luminosity and \teff\ -- to match the star's position in the H-R diagram. If its iron content is equal to $\sim-2.4$~dex or lower, there are no current sets of stellar models able to match its luminosity and \teff. 

A possibility to improve the agreement of BaSTI (as well as Dartmouth and PARSEC) models with the star H-R diagram 
is to use different BCs in the model calculations, or change the mixing length while keeping the choice of the BCs fixed.  
For instance either the use of the HM74 atmospheric thermal stratification to calculate the BCs, or a significant reduction of the mixing length by about 0.2 with respect to the solar calibrated value, could reconcile the models with 
the position of the star in the H-R diagram (within the errors).

Observationally, a more stringent test of the metal poor 
RGB models' \teff\ scale requires not only a more robust assessment of the observational properties of HD~122563 via a more accurate astrometric, spectroscopic and interferometric studies, but also a larger sample of metal-poor stars with accurate empirical determinations of their 
luminosities, \teff, chemical composition and -- if possible -- mass, using asteroseismology
in the case of single stars, or dynamical measurements in the case of binary systems.

From a theoretical point of view, it is crucial to minimise the uncertainties in the calculation of 
the superadiabatic temperature gradients in the convective 
envelopes of RGB models, and the calculation of the outer boundary conditions. The current 
generation of 3D radiation-hydrodynamics simulations of stellar convective envelopes \citep[e.g.,][]{m13, trampedach14, t14b, mwa15} reach the metallicity regime 
of HD~122563 \citep{m13}, but do not cover 
surface gravities low enough to be employed to help model stars like HD~122563. An extension of this type of simulations to lower surface gravities is much needed.

\begin{acknowledgements}
SC has been funded by the European Union – NextGenerationEU" RRF M4C2 1.1  n: 2022HY2NSX. "CHRONOS: adjusting the clock(s) to unveil the CHRONO-chemo-dynamical Structure of the Galaxy” (PI: S. Cassisi), and INAF 2023 Theory grant "Lasting" (PI: S. Cassisi). SC also warmly acknowledges the kind hospitality at the Universit{\'e} de la C{\^o}te d'Azur, Nice, where a large part of this  investigation has been performed. MS acknowledges support from The Science and Technology Facilities
Council Consolidated Grant ST/V00087X/1. 
The authors acknowledge the financial support from the scientific council of the Observatoire de la Côte d’Azur under its strategic scientific program. SC warmly thanks D. Vandenberg for providing his stellar evolution models as well as for several interesting discussions.
This work has made use of data from the European Space Agency (ESA) mission
{\it Gaia} (\url{https://www.cosmos.esa.int/gaia}), processed by the {\it Gaia}
Data Processing and Analysis Consortium (DPAC,
\url{https://www.cosmos.esa.int/web/gaia/dpac/consortium}). Funding for the DPAC
has been provided by national institutions, in particular the institutions
participating in the {\it Gaia} Multilateral Agreement.

\end{acknowledgements}

%-------------------------------------------------------------------
\bibliographystyle{aa}
\bibliography{halo_giant}

\end{document}